\documentstyle[11pt,flushrt,aaspp4]{article}
\newcommand{\et}{et al.}
\newcommand{\rxte}{{\it RXTE}}
\newcommand{\xte}{{\it RXTE}}
\newcommand{\exosat}{{\it EXOSAT}}
\newcommand{\asca}{{\it ASCA}}

\newcommand{\sigratio}{\sigma^{2}_{1{\rm d}}/\sigma^{2}_{300{\rm d}}}

\slugcomment{}
\lefthead{Markowitz \& Edelson}
\righthead{Long-term X-ray Variability of Seyfert~1 Galaxies}

\begin{document}
\title{AN {\it RXTE} SURVEY OF LONG-TERM X-RAY VARIABILITY IN SEYFERT~1 GALAXIES}

\author{A. Markowitz $\&$ R. Edelson}
\affil{{\footnotesize agm,rae@astro.ucla.edu}\\ U.C.L.A. Department of Astronomy;
       Los Angeles, CA 90095-1562; USA}

\begin{abstract}

Data from the first three years of \xte\ observations have been systematically
analyzed to yield a set of 300~day, 2-10~keV light curves with 
similarly uniform, $\sim$5~day sampling, for a total of nine 
Seyfert~1 galaxies. This is the first X-ray variability survey
to consistently probe time scales longer than a few days in a large number
of AGN. Comparison with \asca\ data covering a similar 
band but much shorter ($\lesssim$1 day) time scales shows that all the AGN
are more strongly variable on long time scales than on short time scales. 
This increase is greatest for the highest-luminosity sources.
The well-known anticorrelation between source luminosity and variability
amplitude is both stronger
% in correlation strength 
and shallower in power-law slope when measured on long time scales.
This is consistent with a picture in which the X-ray variability of
Seyfert~1s can be can be described by a single, universal fluctuation
power density shape for which the cutoff moves to longer time scales for
higher luminosity sources.
All of the Seyfert~1s exhibit stronger variability in the
relatively soft 2-4~keV band than in the harder 
7-10~keV band. This effect is much too pronounced
to be explained by simple models based on either the dilution of the
power-law continuum by the Compton reflection component or on
the hard X-rays being produced in a static, pair-dominated,
plane-parallel Comptonizing corona.

\end{abstract}


\section{ Introduction }

X-ray observations can provide constraints on physical conditions in the
innermost regions of Active Galactic Nuclei (AGN), as the X-rays are
generally thought to originate very close to the central engines.
On the basis of spectroscopic observations, the leading models of the
X-ray continuum production are generally based on a hot, Comptonizing
electron or electron-positron pair corona above the accretion disk which
multiply-upscatters thermal soft photons from the disk to produce an X-ray
power-law in the energy range 1-100~keV (e.g., Haardt, Maraschi \&
Ghisellini 1994). Furthermore, the disk, or some other cold, optically 
thick material, reprocesses the hard X-rays, as evidenced by the so-called 
'Compton reflection humps' above $\sim$10~keV in Seyfert~1 spectra, as well as 
strong iron fluorescent lines at $\sim$6.4~keV
(Lightman \& White 1988, Guilbert \& Rees 1988, Pounds \et\ 1990).
AGN also exhibit rapid, aperiodic variability, for which no fully satisfying
explanation has been advanced.
However, at least on short time scales, the variability amplitude has long
been known to correlate inversely with the source luminosity for normal
Seyfert~1s (e.g., Barr \& Mushotzky 1986, Lawrence \& Papadakis 1993, 
Nandra \et\ 1997).

Probably the best way to characterize AGN variability, if
adequate data exist, is to measure the fluctuation power density spectra (PDS).
On short time scales ($\lesssim$1~d), \exosat\ showed that AGN PDS are 
well-described as power-laws ($ P \propto f^s $, where $P$ is the power at
temporal frequency $f$), with index $ s = -1 $ to $-2$ (e.g., Green,
McHardy \& Letho 1993, Lawrence \& Papadakis 1993).
However, only the {\it Rossi X-ray Timing Explorer} (\rxte), with its
combination of flexible scheduling, relaxed pointing constraints, high
throughput and rapid slew speed, has been able to monitor AGN variations
over long time scales.
Only one Seyfert~1 galaxy, NGC~3516, has been subjected to a combination
of even sampling on long, medium and short time scales needed to produce
the first AGN PDS to span time scales of minutes to months.
This PDS showed a flattening on a time scale of $\sim$1~month (Edelson \&
Nandra 1999).
Unfortunately, the \rxte\ archives do not currently contain adequate data
to measure such detailed PDS for other AGN, although it is expected that the
current round of observations will make this feat 
possible within a few years.  

However, the archived data can be used for a less detailed study of the
long-term continuum variability properties of Seyfert 1s.
This paper presents the first such study, 
a survey of homogeneously-sampled, $\sim$300 day light
curves of nine Seyfert~1s observed by \rxte.
The data reduction and sampling are described in \S~2 and the analysis is
described in \S~3.
As discussed in \S~4, 
this has allowed significant insight into AGN
variability, such as quantifying 
the relation between source luminosity and PDS
cutoff frequency.
Furthermore, the long-term spectral variability 
properties can be used to test spectral formation models.
A short summary is given in \S~5. 

\section{ Data Collection and Reduction }

\xte\ observed $\sim$105 AGN for a total 
of $\sim$13~Msec during the first 3 years of its
mission (calibration phase and Cycles~1-3).
Almost all of these data had become public by the time that 
these analyses were performed
(February 2000). This paper utilized these data as well as the
authors' proprietary observations of five Seyfert~1s observed during Cycle~4. 
Section~2.1 describes the reduction of 
the {\rxte} data set used in this paper. 

The goal of this project
was to obtain long time scale monitoring sampled as uniformly 
as possible for as large a number of objects as possible. 
Optimizing this tradeoff
for the currently available archive yielded a sample of nine Seyfert~1s
that were sampled an average of once every 5--13 days for a total period of
300 days, along with adequate short time scale {\asca} data. The 
construction of this sample is described 
in $\S$2.2. The {\asca} data and analysis are discussed in 
$\S$2.3.

\subsection { {\it RXTE} Data }

The {\it RXTE}'s Proportional Counter Array (PCA)
consists of five identical proportional counter units (PCUs; see Swank 1998).
For simplicity and uniformity, data were collected only from 
those PCUs that did not
suffer from repeated breakdown during on-source time
(PCUs 0 and 2 for the entire data set, plus PCU 1 for observations 
prior to 1998 December). 
All quoted count rates for this paper are normalized to 1~PCU. The data were 
reduced using standard extraction methods and {\sc FTOOLS~v4.2} 
software. Data were rejected if they were 
gathered less than 10$\arcdeg$ from the Earth's limb, if 
they were obtained within 30~min after the satellite's passage 
through the South Atlantic Anomaly (SAA), if {\sc ELECTRON0} $>$ 0.1, 
or if the satellite's pointing offset was greater 
than 0$\fdg$02. 

As the PCA has no simultaneous background monitoring capability, 
background data were estimated 
by using {\sc pcabackest~v2.1b} to generate model files
based on the particle-induced 
background, SAA activity, and the diffuse X-ray background. 
This background subtraction is the
dominant source of systematic error in \xte\ AGN monitoring data.
The background models are best calibrated for the topmost PCU layer,
so only counts extracted from this layer were used; 
restricting to the topmost layer
also maximized the signal-to-noise ratio.
All of the targets were faint ($<$~40~ct~s$^{-1}$~PCU$^{-1}$),
so the applicable 'L7-240' background models were used. 
Because the PCU gain settings changed
three times since launch, the count rates were rescaled to a common 
gain epoch. 
Light curves binned to 16~s were generated for all targets over the 
2-10~keV bandpass, where the PCA is most sensitive and the 
systematic errors and background are
best quantified. Light curves were also generated 
for the 2-4~keV and 7-10~keV bands.
The data were then binned on orbital time scales; orbits with less than 
10 points were rejected. 
Errors on each point were obtained from the standard deviations
of the data in each orbital bin.
Further details of \xte\ data reduction can
be found in, e.g., Edelson \& Nandra (1999).

\subsection{ Optimized Sampling and Source Selection }

The main goal of this project was to understand the
long-term X-ray variability properties of Seyfert~1s.
This required producing a substantial sample of 
objects that were (to the greatest degree possible)
uniformly monitored on long time scales for good comparison between sources. 
Furthermore, there was required to exist adequate short-term data for 
comparison of the two time scales.
Objects were also required to contain
at least 20 points in the final light curve.
Sources with a weighted mean count rate significantly below
1~ct~s$^{-1}$~PCU$^{-1}$ over the full 2-10~keV 
bandpass were rejected to minimize the risk of contamination from 
faint sources in the field-of-view and to ensure adequate signal-to-noise.
%It was necessary to utilize
%an optimum window size (300~d was chosen) and resample the light curves
%at as similar a rate as possible, as described below.
%The availability of adequate short \asca\ data was required in order to
%allow comparison of long and short time scale variability, as discussed later.

The sampling of the publically available Cycle~1-3 data
was highly uneven in general, as observations were made 
with a wide variety of science goals. The resulting light curves 
featured a wide range of sampling patterns and durations.
Targets with observations
that spanned less than the optimum window of $\sim$300~d were rejected. 
Choosing a longer window would have resulted in too few sources
in the final survey, 
and a shorter window would have been less 
appropriate for long-term variability analysis,
as many sources had $\sim$2~month gaps of 1-year 
time scales due to satellite $\beta$-angle viewing constraints. 
Targets with gaps
greater than 33$\%$ of the total duration were also
rejected, as gaps reduce the statistical significance of 
parameters derived over the full duration. 
For each target, the most densely sampled 300~d that did not include
significant gaps was selected for analysis.
 
To extract a data set that was as uniformly sampled as possible, 
it was necessary to resample at a common, optimized sampling rate.
This was done with an
algorithm that kept spaces between adjacent accepted points
as close to 5~d as the original sampling pattern allowed.
Resampling at a rate longer than 5~d would have
resulted in too few points in the final light curves, while resampling 
significantly more frequently would have yielded 
light curves that were not sufficiently uniformly sampled
a data set given the original range of sampling patterns.

This reduction yielded a sample of fifteen AGN, including
two Seyfert~2s (NGC~4258 and NGC~5506) and four blazars
(3C~273, 3C~279, 3C~454.3 and PKS~1510-089) that 
will not be considered further in this paper.
Excluding those objects yielded a sample of nine 
Seyfert~1s (3C~120, Ark~120, Fairall~9, 
MCG--6-30-15, NGC~3516, NGC~3783, 
NGC~4051, NGC~4151, and NGC~5548). 
Figure~1 shows the full 2-10~keV light curves for these Seyfert~1s, ranked by
source luminosity.
Figure~2 shows the light curves after applying the $\sim$300~d 
window and resampling.
Table~1 summarizes the source observation and sampling
parameters for the survey targets
over the 2-10~keV bandpass. All source fluxes were calculated 
using HEASARC's online W3PIMMS~v3.0
Flux Converter assuming an intrinsic power-law with a photon
index obtained from the online Tartarus archive of \asca\ AGN observations 
(http://tartarus.gsfc.nasa.gov; e.g.\ Nandra \et\ 1997, 
Turner \et\ 1999). 
%Galactic absorption was taken into account (Dickey \& Lockman 1990).
Luminosities were calculated assuming
$H_{o}$~=~75~km~s$^{-1}$~Mpc$^{-1}$ and $q_{o}$~=~0.5.

%\tablehere{1}

%\figurehere{1} 

%\figurehere{2} 

\subsection { {\it ASCA} Data }

Short-term 2-10~keV light curves were also obtained
from the Tartarus database. 
The count rates in the light curves provided had been
combined and averaged between {\it ASCA}'s two
Solid-state Imaging Spectrometers (SIS; Burke \et\ 1994, Gendreau 1995) and
binned to 16~s.
For each target, the observation with the
longest duration available in the archive was selected,
and only the first 1~day (maximum duration of 86.4~ks) 
was used for these analyses.
The light curves were rebinned on orbital time scales with 
the same algorithm used for the \rxte\ data to yield light curves
that were generally 10-15 consecutive orbital bins.  
Background light curves were similarly binned and subtracted to
produce net count rate light curves.
Table~2 lists the source observation and sampling 
parameters for the \asca\ data.  

%\tablehere{2}

\section{ Analysis }

\subsection{ Excess Variance }

The normalized excess variance, $\sigma^{2}_{RMS}$, 
was utilized to quantify the amplitude of variability 
in each light curve (e.g., Nandra \et\ 1997):
\[\sigma^{2}_{RMS} = \frac{1}{{\rm N}\mu^{2}}\sum_{i=1}^{{\rm N}} [(X_{i}-\mu)^{2}-\sigma_{i}^{2}] \] 
where the count rates for 
the N~points in each light curve are $X_{i}$, with errors 
$\sigma_{i}$, and 
$\mu$ is the unweighted arithmetic mean.
Table~3 lists the excess variances for each target over 
the 2-4~keV band, 7-10~keV band, and full 2-10~keV bandpass
for the \xte\ data, 
the ratio of the 300~d 2-4~keV to 7-10~keV variances, and excess variances
over the full 2-10~keV bandpass for the \asca\ data,.
($\sigma^{2}_{{\rm 300d,soft}}$, $\sigma^{2}_{{\rm 300d,hard}}$, 
$\sigma^{2}_{{\rm 300d}}$, 
$\sigma^{2}_{{\rm soft}}$$/$$\sigma^{2}_{{\rm hard}}$, 
and $\sigma^{2}_{{\rm 1d}}$, respectively).

%\tablehere{3}

\subsection{ Construction of Correlation Diagrams }

Figure~3 displays the excess variance for both the
long (300~day) and short (1~day) time scale data
plotted against source luminosity over the 2-10~keV bandpass. 
For all sources, the long-term excess variances are greater than the 
short-term excess variances.
Both data sets conform well to a power-law of the form 
$\sigma^{2}_{RMS}$~$\propto$~$L_{x}^{-a}$,
an anticorrelation observed before in AGN (e.g.\ 
Green, McHardy, \& Lehto 1993, Nandra \et\ 1997, Turner \et\ 1999).
The nine short-term data points were fitted by a logarithmic 
slope $a$~=~0.861~$\pm$~0.070. 
The long-term data are described by a power-law with a
much shallower slope, $a$~=~0.277~$\pm$~0.009.
The correlation is much stronger for the long-term data
(with a correlation coefficient of r~=~$-$0.972) than for 
the short-term data (where r~=~$-$0.839). 

%\figurehere{3}

The zero-lag correlation between $\sigma^{2}_{{\rm 300d,soft}}$ and 
$\sigma^{2}_{{\rm 300d,hard}}$
is shown in Figure~4. All of the sources exhibit 
stronger variability in the relatively softer X-rays than in the harder X-rays,
as the slope of the best-fit line is 0.724~$\pm$~0.041,
indicating that the soft X-rays make an increasingly dominant contribution
to the overall variability for the more variable sources.

%\figurehere{4}

Another test of spectral variability is to plot
the 7-10~keV count rates against the 
2-4~keV count rates for individual sources (see Figure~5).  
The resulting spectral variability slopes, listed in Table~3, 
range from 1.14 to 1.81, 
again indicating that the relatively soft band is more strongly
variable.

%\figurehere{5}

Figure~6 plots this spectral variability slope
against source luminosity and $\sigma^{2}_{{\rm 300d}}$.
There is weak evidence for an anticorrelation
between the spectral variability slope and source luminosity in the sense that
less luminous sources exhibit more variability in the soft band.
There is weak evidence for a positive correlation
between between the spectral variability slope and 
2-10~keV $\sigma^{2}_{{\rm 300d}}$. 

%\figurehere{6}

Table~4 summarizes the correlations 
discussed in this section.

%\tablehere{4}

\section{ Discussion }

On long time scales,  
the anticorrelation between source luminosity and variability is 
stronger (higher correlation coefficient), yet shallower in power-law 
slope, compared to short time scales.
The shallower slope on long time scales can be viewed as the highest
luminosity sources showing the greatest
increase in long-term variability amplitude.
As discussed below, this trend
can be explained by a simple scaling of PDS turnover
frequency with luminosity.
Furthermore, all of the targets exhibit stronger variability at relatively
softer X-ray energies (2--4~keV) than at harder 
X-ray energies (7--10~keV), which
allows tests of simple X-ray
reprocessing models.

\subsection{ Variability-Luminosity Relationship and PDS Movement }

As mentioned in the introduction, only one AGN (NGC~3516) currently has
adequate data to clearly measure a cutoff in the PDS (at a time scale of
$\sim$1~month; Edelson \& Nandra 1999).
A toy model can be constrained assuming
that all AGN PDS have a universal
shape, with the location of the break (in both amplitude and time
scale) changing with luminosity.
The observed anticorrelation between source luminosity and variability
amplitude observed on short time scales (e.g., Lawrence \& Papadakis
1993; Nandra \et\ 1997) could then
be due either to a positive correlation
between luminosity and cutoff time scale or to
an inverse correlation between
luminosity and the overall variability amplitude.
This is illustrated in Figure~7, with the first case of
changing time scale shown on the left (e.g., less luminous sources vary
more rapidly) and the second case of changing amplitude on the right
(e.g., less luminous sources vary more strongly).
It is also illustrated in Figure~8, except in $ f*P_f -f $ space instead
of $ P_f -f $ space.

Within this model, it is possible to use the current data to discriminate
between these two pictures.
If the amplitude scales with luminosity but is independent of time scale,
then the
ratio of short- to long-term variances, $\sigratio$, will remain independent
of luminosity.
However, if the time scale changes with luminosity, then $\sigratio$ will be
larger for more luminous sources.
Figure~9 is such a plot of $\sigratio$ as a function of luminosity, with
the solid line derived by scaling the measured PDS of NGC~3516 (with a
high-frequency slope of $-$1.76 and turnover 
frequency of 4~$\times$~10$^{-7}$~Hz) linearly
with luminosity in  $ P_f -f $ space.
The short-term variances were obtained by integrating between
temporal freqencies of 1~d$^{-1}$
and 15~d$^{-1}$ (one satellite orbit).
The long-term variances were calculated
using the frequency range 0.003~d$^{-1}$ to 15~d$^{-1}$.
%(The upper limit is not 0.2~d$^{-1}$ since data points in the resampled light curves are not 5~d averages but rather data points selected to be adequately separated.)
No arbitrary scaling was done.
This result is roughly consistent with the idea that the variability time
scale (the duration required for the source to achieve a given level of
r.m.s.\ variability) is in proportion to the size of the emitting region
in AGN and hence black hole mass, which in turn determines luminosity.

Finally, Figure~3 also shows that the long-term correlation is much
stronger (more highly correlated) than the short-term correlation.
This could be due to the a simple data effect: the long-term variability
is stronger the short-term, and \rxte\ gets higher signal-to-noise ratios
than \asca.
This would result in a higher variability-to-noise ratio for the long-term
data, and therefore a stronger correlation.

\subsection{ Spectral Variability results }

Figures~4 and 5 demonstrate that all sources in the sample show stronger
variability in the 2-4~keV band compared to the 7-10 keV band.
This trend has been observed previously in Seyfert~1s, e.g., by Turner,
George, \& Nandra (1998).
Green, McHardy, \& Lehto (1993) model such spectral softening as the
result of a constant reflection component superimposed on a variable soft
component.
However, this model is not consistent with the data reported in this
paper.
For the spectral model {\sc pexrav} in XSPEC~v11.00, 
consisting of an absorbed power-law plus Compton
reflection component (Magdziarz \&
Zdziarski, 1995), the reflected component accounts for at most $\sim$8$\%$
of the total 7-10~keV flux.
Assuming this maximum effect, a constant reflection component can reduce
$\sigma^{2}_{{\rm 300d,hard}}$ by as much as $\sim$18$\%$.
However, for all objects in the present sample, $\sigma^{2}_{{\rm
soft}}$/$\sigma^{2}_{{\rm hard}}$ is still larger than 1.18; as seen in
Table~3, these values range from 1.30 to 3.85.
This result suggests that the observed spectral variability may instead be
due to changes in the intrinsic spectral slope $\Gamma$.

Haardt, Maraschi, \& Ghisellini (1997) discuss a model for Comptonizing
coronae in which optical depth variations in a fixed-geometry,
plane-parallel, pair-dominated, optically thin
(0.1~$\le$~$\tau$~$\le$~1.0) corona drive spectral changes in the
Comptonized emission in that spectral steepening and a decrease
in corona temperature accompany increases in optical depth.
The model predicts that a steepening of $\Gamma$ by 0.2 over the 2-10~keV
bandpass corresponds to an increase in 2-10~keV flux by a factor of 10
(valid for $\Gamma$~$\lesssim$~2.0).
Assuming that the Comptonized power-law pivots at an energy of 10~keV, the
2-4~keV flux is therefore expected to increase by 25$\%$ more than the
7-10~keV flux increases.
However, this model also is inconsistent with the data.
The average increase of 2-10~keV flux for all targets in the present
sample from the lowest to highest flux states is a factor of
$\sim$4; scaling the model's prediction accordingly implies
$\Delta$$\Gamma$~=~+0.12, and hence the 2-4~keV flux should increase by
14$\%$ relative to the 7-10~keV flux.
For the present sample, however, the average increase is 53$\%$
(see Figure~5).
There is more spectral variability and less total
(2-10~keV) flux variability than the model predicts.
Reasons for this discrepancy include the possibility that pairs may
not dominate in coronae.
However, Haardt, Maraschi, \& Ghisellini (1997) also find that for
a pair-dominated corona in which variations of the scale size of
the corona are important and intrinsic flux variations are negligible,
significant spectral variations could be expected without large
variations in the 2-10~keV flux.
This situation, which could arise from a corona composed 
of many active regions
instead of a single homogeneous region, is more consistent with the present
observations.

\section{ Conclusions }

This paper reports the first X-ray variability survey 
with highly uniform and consistent sampling to
probe long time scales ($\sim$300~d) in a reasonably large sample of
Seyfert~1 galaxies. The variability amplitudes were found to be greater
on long time scales than on short time scales, with the largest increases
seen in the most luminous sources. This resulted in the slope
of the anticorrelation between excess variance and luminosity
being shallower on long time scales.
This trend can be explained if the time scale for a variability mechanism
increases for more luminous sources, as would be the case if
all Seyfert~1s radiate
at approximately the same ratio of L$/$L$_{{\rm Edd}}$.
Consequently, a PDS corresponding to a higher luminosity object will be
displaced towards lower temporal frequencies relative to that of a lower
luminosity object.

All of the Seyfert~1s exhibited 
stronger variability in the relatively soft 2-4~keV
band than in the harder 7-10~keV band, too strong
to be explained by a simple model based
on the dilution of the power-law continuum in the 7-10~keV band by a
constant Compton reflection component.
The data are also inconsistent with a simple model of a static,
pair-dominated, plane-parallel Comptonizing corona.

Recently, systematic monitoring with \xte\ has been started for several
Seyfert~1s, with even sampling over a wide range of time scales in the
same pattern as the NGC~3516 campaign.
When these campaigns are completed in 2002, it should be possible to measure
the PDS over four decades of temporal frequency, hopefully allowing direct
measurement of the turnover and determination of how PDS shape
evolves with luminosity, as predicted in this paper.

\acknowledgments 
The authors would like to thank Tess Jaffe and Simon
Vaughan for help with data analysis.
This work has made use of data obtained through the High Energy
Astrophysics Science Archive Research Center Online Service, provided by
the NASA Goddard Space Flight Center, the TARTARUS database, which is
supported by Jane Turner and Kirpal Nandra under NASA grants NAG~5-7385
and NAG~5-7067, and the NASA$/$IPAC Extragalactic Database which is
operated by the Jet Propulsion Laboratory, California Institute of
Technology, under contract with the National Aeronautics and Space
Administration.
The authors acknowledge financial support from NASA grants NAG~5-7315 and
NAG~5-9023.

%\newpage

\newpage

\begin{figure}[ht]
\epsscale{0.93}
\plotone{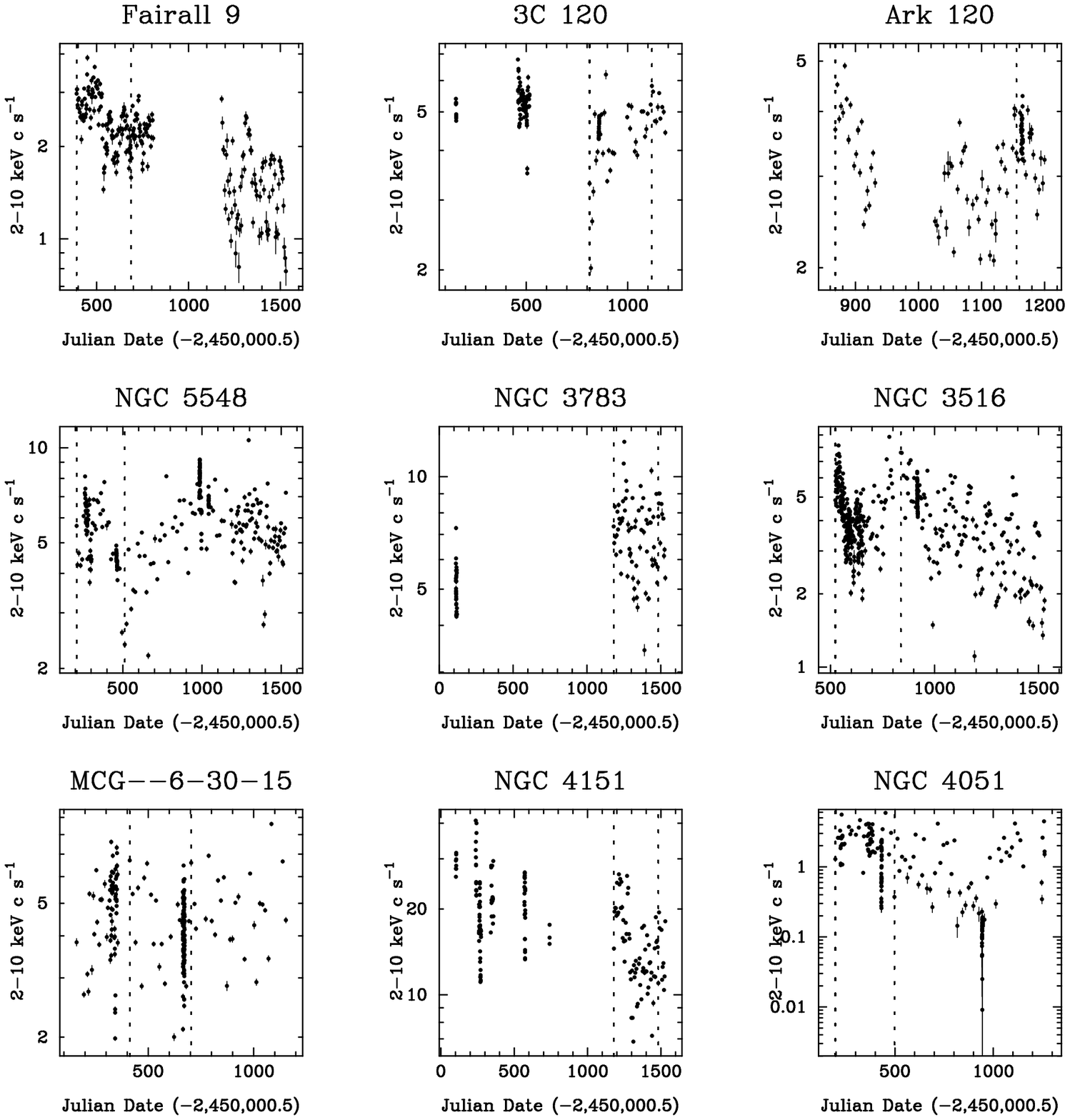}
\figcaption{\rxte\ 2-10~keV light curves, ranked by luminosity,
before applying the $\sim$300~d window and subsampling.
Error bars are 1$\sigma$. The dashed vertical lines 
indicate the $\sim$300~d period used for
analysis.}
\label{fig1}
\end{figure}

\begin{figure}[ht]
\epsscale{0.93}
\plotone{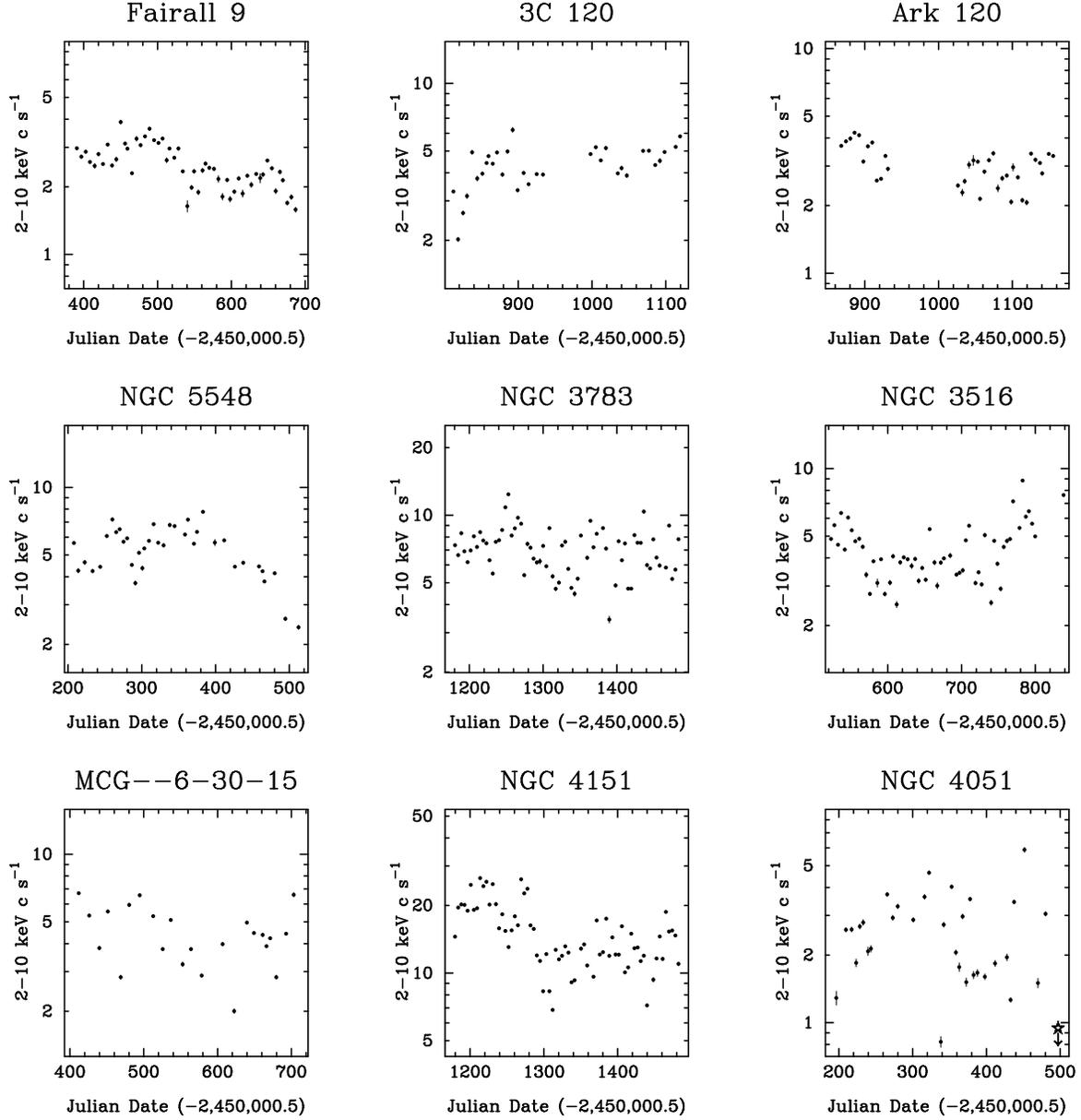}
\figcaption{\rxte\ 2-10~keV light curves, ranked by luminosity,
after applying the $\sim$300~d window and 
subsampling to 5~d. Error bars are 1$\sigma$. 
The y-axis has been scaled such that the
maximum and minimum values
are, respectively, 3.5 and $(3.5)^{-1}$~$\times$ the
mean count rate. A data point in NGC~4051 corresponding
to JD$-$2,450,000.5~=~497.30, rate~=~0.37~$\pm$~0.03 has been omitted.}
\label{fig2}
\end{figure}

\begin{figure}[ht]
\epsscale{0.65}
\plotone{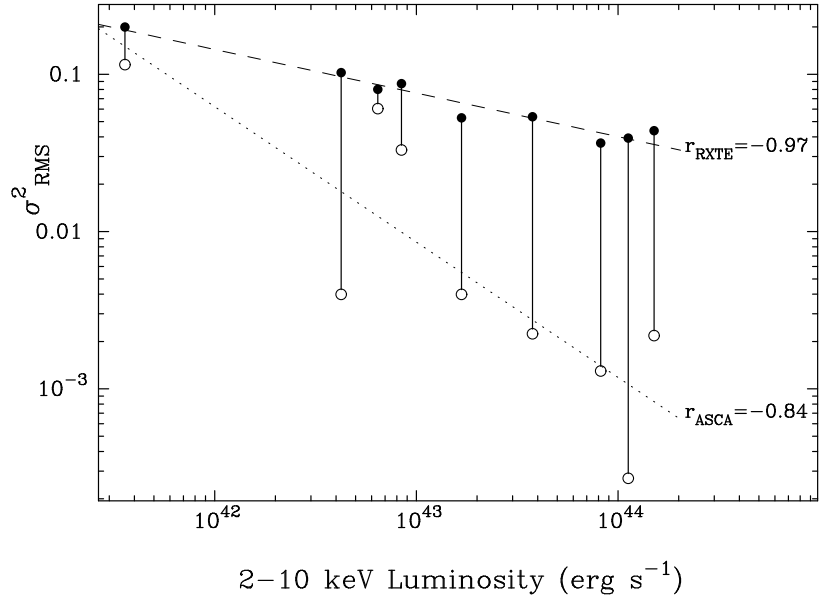}
\figcaption{\rxte\ long-term excess variance and \asca\ short-term
excess variance plotted against source luminosity over
the 2-10~keV bandpass. Filled circles, best-fit by the dashed line, 
are \xte\ data; open circles, best-fit by the dotted line,
are \asca\ data. Note the tighter correlation in the \xte\ data points.}
\label{fig3}
\end{figure}

\begin{figure}[ht]
\epsscale{0.50}
\plotone{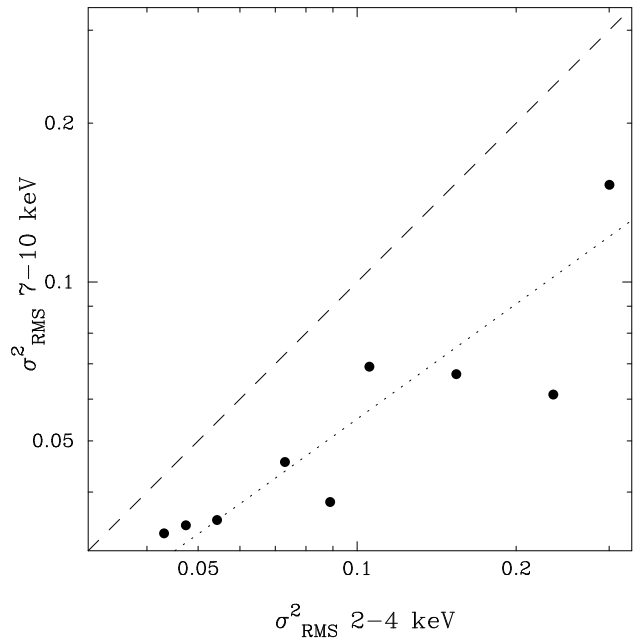}
\figcaption{\rxte\ hard X-ray excess variance plotted against
soft X-ray excess variance. A source with equally strong variability
in the hard and soft bands would lie on the dotted line.
All of the sources exhibit stronger variability in
softer X-rays than in hard X-rays.}
\label{fig4}
\end{figure}

\begin{figure}[ht]
\epsscale{0.85}
\plotone{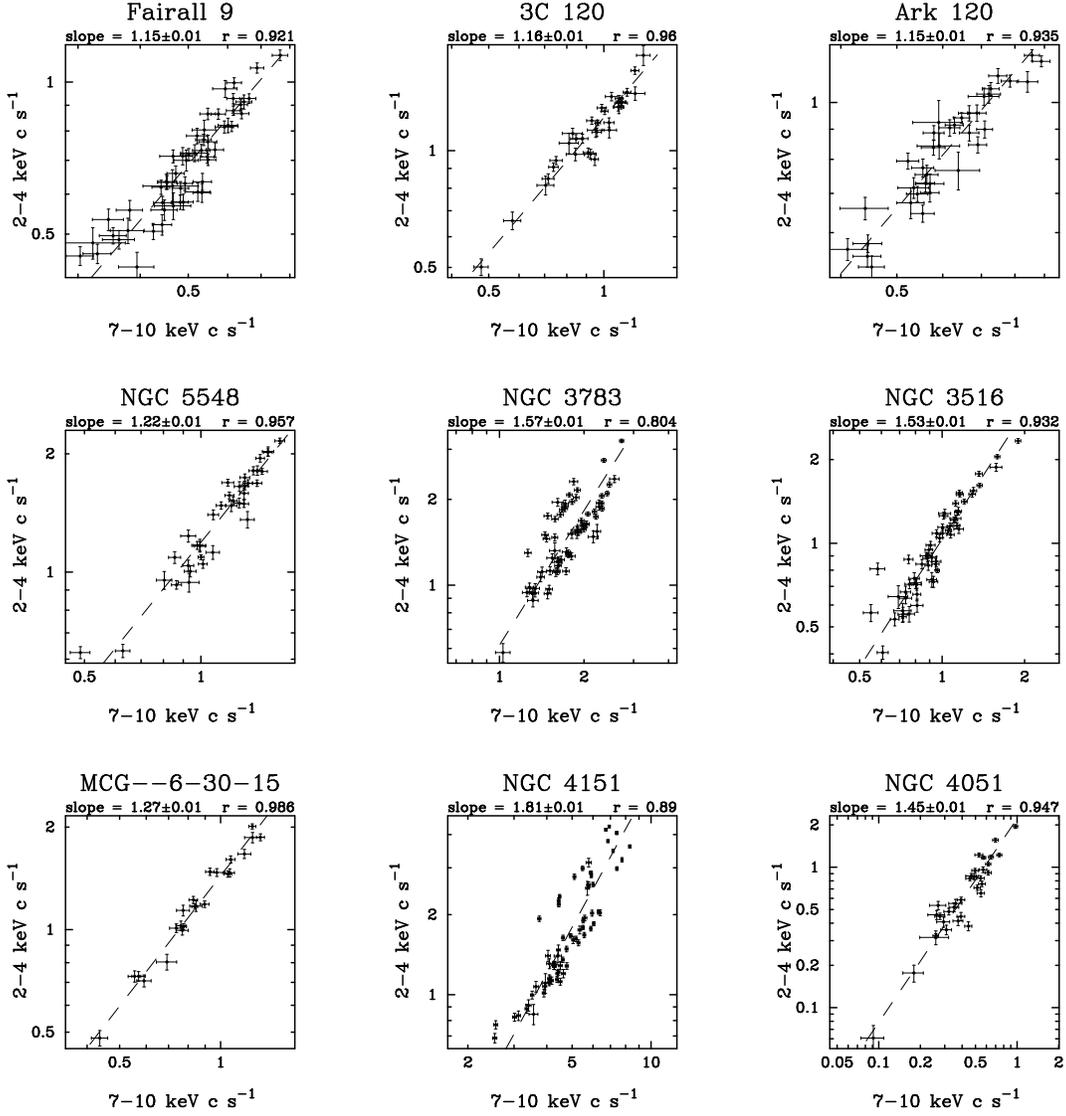}
\figcaption{\rxte\ 2-4~keV count rates plotted against 7-10~keV count rates
for the clipped and subsampled light curves, in order of descending
2-10~keV source luminosity. The resulting
slope, described by the best-fit dashed lines, 
is the spectral variability slope.
r is the correlation coefficient for each plot.}
\label{fig5}
\end{figure}

\begin{figure}[ht]
\epsscale{0.70}
\plotone{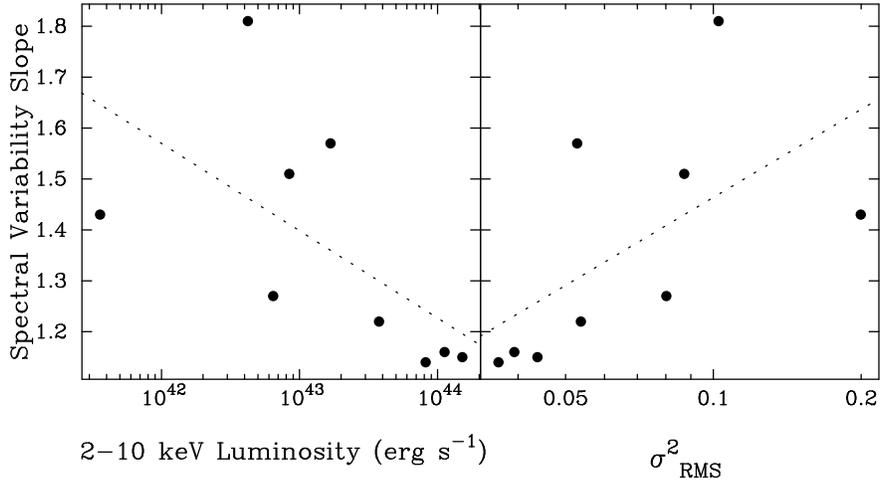}
\figcaption{Spectral variability slope plotted against
2-10~keV source luminosity (left) and \xte\ 2-10~keV 
long-term excess variance (right). In each plot, the dotted line
indicates the best-fit line.}
\label{fig6}
\end{figure}

\begin{figure}[ht]
\epsscale{0.70}
\plotone{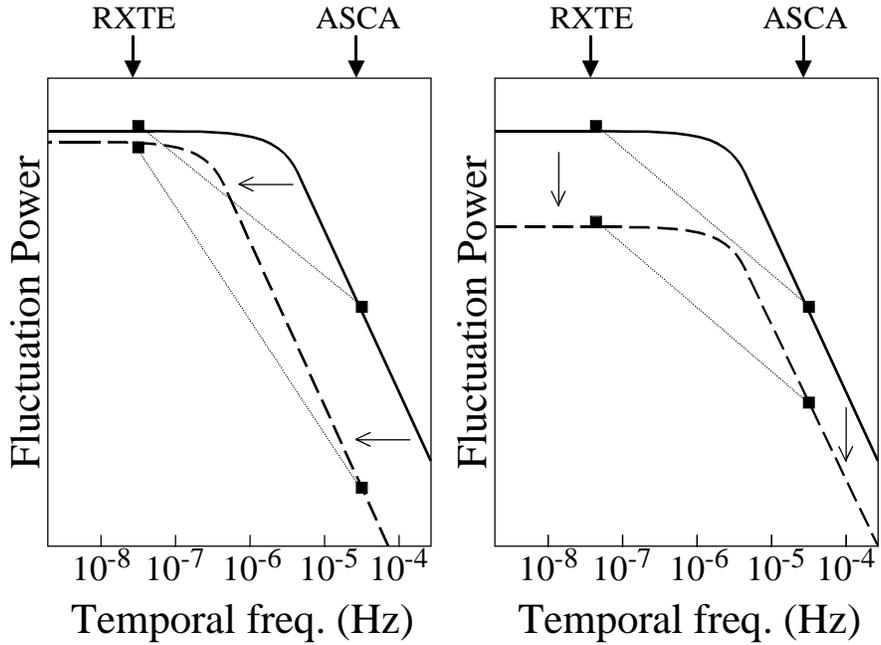}
\figcaption{Model of a characteristic
PDS change with increasing source luminosity, plotted
in  $ P_f -f $ space; the
dashed-line PDS corresponds to a source with higher luminosity than 
the source corresponding to the solid-line PDS. If the PDS
moves to the left in temporal frequency (left), then the 
ratio of low-frequency variability to
high-frequency variability will be greater than in lower luminosity
sources. If the PDS moves downward
in fluctuation power (right), then the ratios of low- to high-frequency
variability should be equal and independent of luminosity.} 
\label{fig7}
\end{figure}

\begin{figure}[ht]
\epsscale{0.69}
\plotone{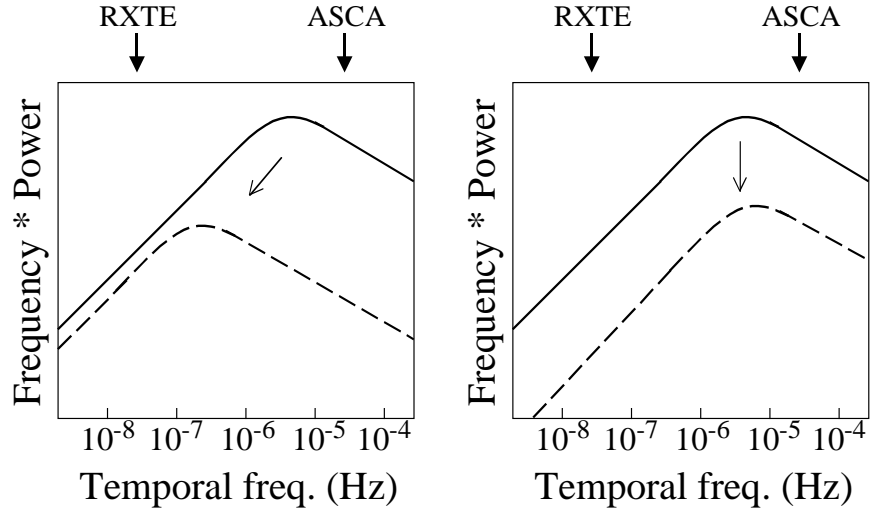}
\figcaption{Model of a characteristic
PDS change with increasing source luminosity, plotted in $ f*P_f -f $
space. The
dashed-line PDS corresponds to a source with higher luminosity than 
the source corresponding to the solid-line PDS.} 
\label{fig8}
\end{figure}

\begin{figure}[ht]
\epsscale{0.71}
\plotone{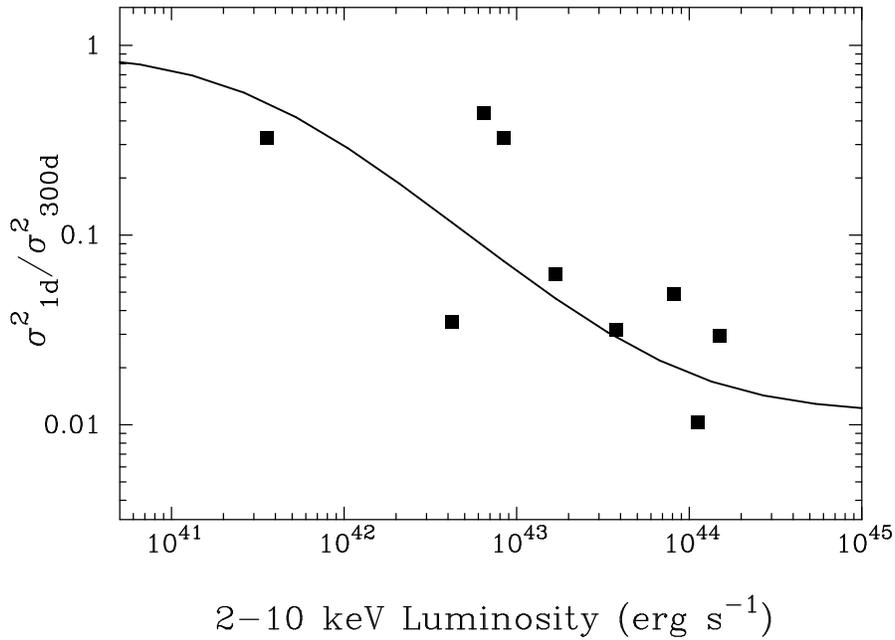}
\figcaption{Predicted and measured values of $\sigratio$
(solid line and squares, respectively)
plotted against luminosity. No arbitrary scaling was
done.}  
\label{fig9}
\end{figure}

\newpage

\begin{deluxetable}{ccccccc}
\tablewidth{0pc}
\tablenum{1}
\tablecaption{Source and \rxte\ Sampling Parameters \label{tab1}}
\small
\tablehead{
\colhead{} & \colhead{} & \colhead{} & \colhead{} & \colhead{Mean} &
\colhead{} & \colhead{Lumin.} \\
\colhead{} & \colhead{} & \colhead{Num.} & \colhead{JD-2450000.5} & \colhead{Count Rate} & 
\colhead{Mean} & \colhead{log($L_{{\rm 2-10~keV}}$} \\
\colhead{Name} & \colhead{z} & \colhead{Points} & \colhead{Range} &
\colhead{(ct s$^{-1}$)} & \colhead{S$/$N} & \colhead{{\footnotesize (erg s$^{-1}$))}} }
\startdata
Fairall 9     & 0.047 & 54 & 50390.6-50687.2 & 2.5  & 56 & 44.18\nl 
3C~120        & 0.033 & 33 & 50812.1-51119.4 & 4.3  & 73 & 44.05\nl 
Ark 120       & 0.032 & 36 & 50868.1-51155.3 & 3.0  & 56 & 43.91\nl
NGC 5548      & 0.017 & 37 & 50208.1-50512.6 & 5.1  & 77 & 43.58\nl
NGC 3783      & 0.010 & 68 & 51180.6-51483.6 & 7.1  & 84 & 43.22\nl 
NGC 3516      & 0.009 & 57 & 50523.0-50838.3 & 4.4  & 69 & 42.93\nl
MCG--6-30-15  & 0.008 & 23 & 50411.9-50703.3 & 4.3  & 71 & 42.81\nl
NGC 4151      & 0.003 & 69 & 51179.6-51482.5 & 15.2 & 135 & 42.63\nl
NGC 4051      & 0.002 & 33 & 50196.5-50497.3 & 2.4  & 37 & 41.55\nl 
\enddata
\tablecomments{All quantities are for the 2-10~keV bandpass. The targets are
ranked by 2-10~keV luminosity. Redshifts (Column~2) were
obtained from the NED database. Column~4 is the number of points 
in the \rxte\ light curve after clipping to $\sim$300~d and 
subsampling to 5~d. Column~5 is the weighted mean \rxte\ count rate
per PCU.}
\end{deluxetable}

\vspace{3.0cm}
\begin{deluxetable}{ccccccc}
\tablewidth{0pc}
\tablenum{2}
\tablecaption{\asca\ Sampling Parameters \label{tab2}}
\small
\tablehead{
\colhead{} & \colhead{Num.} & \colhead{JD-2450000.5} & \colhead{Sequence} &
\colhead{Count Rate} & \colhead{Mean} & 
\colhead{} \\
\colhead{Name} & \colhead{Points} & \colhead{Range} & 
\colhead{ID No.} &\colhead {(ct s$^{-1}$)} & \colhead{S$/$N} & \colhead{$\Gamma$} }
\startdata
Fairall 9   & 11 & 49700.6-49701.3 & 73011000 & 0.47 & 19 & 2.18\nl
3C 120      & 15 & 49400.6-49401.6 & 71014000 & 0.94 & 35 & 1.97\nl
Ark 120     & 15 & 49624.8-49625.8 & 72000000 & 0.55 & 25 & 2.02\nl
NGC 5548    & 15 & 49195.6-49196.6 & 70018000 & 0.89 & 26 & 1.79\nl
NGC 3783    & 11 & 50278.2-50278.9 & 74054020 & 1.31 & 30 & 1.43\nl
NGC 3516    & 13 & 49444.1-49445.0 & 71007000 & 1.49 & 42 & 1.60\nl
MCG--6-30-15 & 15 & 49556.2-49557.2 & 72013000 & 0.80 & 26 & 1.80\nl
NGC 4151    & 15 & 49847.1-49848.1 & 73019000 & 1.96 & 62 & 0.45\nl
NGC 4051    & 15 & 49510.6-49511.6 & 72001000 & 0.44 & 20 & 2.07\nl
\enddata
\tablecomments{Column~1
is the number of points in the {\it ASCA} light curve
after orbitally binning. Column~4 is the weighted mean \asca\ 
count rate averaged between 
both SIS instruments. The photon indices in 
column~7 were obtained from the Tartarus database.}
\end{deluxetable}

\begin{deluxetable}{ccccccc}
\tablewidth{0pc}
\tablenum{3}
\tablecaption{Derived Variability Parameters \label{tab3}}
\small
\tablehead{
\colhead{} & 
\colhead{$\sigma^{2}_{{\rm 300d,soft}}$} & 
\colhead{$\sigma^{2}_{{\rm 300d,hard}}$} & 
\colhead{$\sigma^{2}_{{\rm 300d}}$} & \colhead{} &
\colhead{$\sigma^{2}_{{\rm 1d}}$} & 
\colhead{Spectral} \\
\colhead{} & \colhead{{\it RXTE}} & \colhead{{\it RXTE}} & 
\colhead{{\it RXTE}} & $\sigma^{2}_{{\rm soft}}$$/$$\sigma^{2}_{{\rm hard}}$ &
\colhead{{\it ASCA}} &  \colhead{Variability} \\  
\colhead{Name} & \colhead{2-4~keV} & \colhead{7-10~keV} &
\colhead{2-10~keV} & \colhead{{\it RXTE}} & \colhead{2-10~keV}  & 
\colhead{Slope} }
\startdata
Fairall 9 &  0.054  &   0.035 &   0.044  &  1.54 & 0.0022   & 1.15\nl
3C~120    &  0.047  &   0.035 &   0.039  &  1.34 & 0.0003   & 1.16\nl
Ark 120   &  0.043  &   0.033 &   0.037  &  1.30 & 0.0013   & 1.14\nl
NGC 5548  &  0.073  &   0.046 &   0.054  &  1.59 & 0.0022   & 1.22\nl
NGC 3783  &  0.089  &   0.038 &   0.053  &  2.34 & 0.0040   & 1.57\nl
NGC 3516  &  0.154  &   0.067 &   0.087  &  2.30 & 0.0330   & 1.51\nl
MCG--6-30-15 & 0.106 &  0.069 &   0.080  &  1.54 & 0.0604   & 1.27\nl
NGC 4151  &  0.235  &   0.061 &   0.103  &  3.85 & 0.0040   & 1.81\nl
NGC 4051  &  0.301  &   0.153 &   0.200  &  1.97 & 0.1153   & 1.43\nl
\enddata
%\tablecomments{}
\end{deluxetable}

\begin{deluxetable}{cccccc}
\tablewidth{0pc}
\tablenum{4}
\tablecaption{Summary of Correlations \label{tab5}}
\small
\tablehead{
\colhead{Figure No.} & \colhead{x-axis} & \colhead{y-axis} & 
\colhead{r} & \colhead{P(r)} & \colhead{Slope} }
\startdata
3 & $\sigma^{2}_{{\rm 300d}}$ & Luminosity  & $-$0.972 & 1.18$\times$$10^{-5}$ & $-$0.277 $\pm$ 0.009\nl
3 & $\sigma^{2}_{{\rm 1d}}$ & Luminosity & $-$0.839 & 4.69$\times$$10^{-3}$ & $-$0.861 $\pm$ 0.070\nl
4 & $\sigma^{2}_{{\rm 300d,hard}}$ & $\sigma^{2}_{{\rm 300d,soft}}$ & +0.887 & 1.43$\times$$10^{-3}$ & +0.724 $\pm$ 0.041\nl
6 & Luminosity &  Spect. Var. Slope &  $-$0.616 & 7.73$\times$$10^{-2}$ & $-$0.17 $\pm$ 0.03\nl
6 & Long-term $\sigma^{2}_{{\rm 300d}}$ & Spect. Var. Slope & +0.586 & 9.73$\times$$10^{-2}$ & +0.57 $\pm$ 0.10\nl
\enddata
\tablecomments{Luminosity is over the 2-10~keV bandpass.
Short-term data taken from {\it ASCA}; all others 
from {\it RXTE}. Column~4 is the correlation coefficient. 
P(r) in column~5 is the probabililty
of obtaining that correlation coefficient by chance.}
\end{deluxetable}

\end{document}